
\input phyzzx
\input psfig
\def\havefigures{n}   

\catcode`\@=12 	

\font\eighti=cmmi8                          \skewchar\eighti='177
\font\eightsy=cmsy8                          \skewchar\eightsy='60
\font\eightsl=cmsl8
\font\eightit=cmti8
\def\noblackbox{\overfullrule=0pt}
\noblackbox
\def\half{{1\over2}}

\def\bold#1{\setbox0=\hbox{$#1$}%
     \kern-.025em\copy0\kern-\wd0
     \kern.05em\copy0\kern-\wd0
     \kern-.025em\raise.0433em\box0 }
\def\unlock{\catcode`@=11} 
\def\lock{\catcode`@=12} 
\def\Buildrel#1\under#2{\mathrel{\mathop{#2}\limits_{#1}}}
\def\llongrarrow{\hbox to 40pt{\rightarrowfill}}

%
 \newtoks\slashfraction
 \slashfraction={.13}
 \def\slash#1{\setbox0\hbox{$ #1 $}
 \setbox0\hbox to \the\slashfraction\wd0{\hss \box0}/\box0 }
 \unlock
 \def\leftrightarrowfill{$\m@th\mathord-\mkern-6mu%
   \cleaders\hbox{$\mkern-2mu\mathord-\mkern-2mu$}\hfill
   \mkern-6mu\mathord\leftrightarrow$}
 \def\overlrarrow#1{\vbox{\ialign{##\crcr
       \leftrightarrowfill\crcr\noalign{\kern-\p@\nointerlineskip}
       $\hfil\displaystyle{#1}\hfil$\crcr}}}
 \lock
%

{\obeyspaces\global\let =\ }   
%
\def\papersize{	\hsize=35pc\vsize=50pc\hoffset=1cm\voffset=1.3cm
           	\pagebottomfiller=0pc
		\skip\footins=\bigskipamount\normalspace}
\def\lettersize{\hsize=6.5in\vsize=8.5in\hoffset=0cm\voffset=1.6cm
   		\pagebottomfiller=\letterbottomskip
		\skip\footins=\smallskipamount \multiply\skip\footins by 3
		\singlespace}
\papers
%
\catcode`\@=11 
\newif\ifletterstyle		
\letterstylefalse 		
\def\letters{\lettersize\letterstyletrue
   \headline=\letterheadline \footline=\letterfootline
   \immediate\openout\labelswrite=\jobname.lab}
\def\iftpub{\afterassignment\iftp@b\toks@}
\def\iftp@b{\edef\n@xt{\Pubnum={UFTP--\the\toks@}}\n@xt}
\let\pubnum=\iftpub
\expandafter\ifx\csname eightrm\endcsname\relax
    \let\eightrm=\ninerm \let\eightbf=\ninebf \fi
\catcode`\@=12 
%



\unlock
\def\eightpoint{\relax
    \textfont0=\eightrm          \scriptfont0=\eightrm
    \scriptscriptfont0=\fiverm
    \def\rm{\fam0 \eightrm \f@ntkey=0 }\relax
    \textfont1=\eighti           \scriptfont1=\eighti
    \scriptscriptfont1=\fivei
    \def\oldstyle{\fam1 \eighti \f@ntkey=1 }\relax
    \textfont2=\eightsy          \scriptfont2=\eightsy
    \scriptscriptfont2=\fivesy
    \textfont3=\tenex          \scriptfont3=\tenex
    \scriptscriptfont3=\tenex
    \def\it{\fam\itfam \eightit \f@ntkey=4 }\textfont\itfam=\eightit
    \def\sl{\fam\slfam \eightsl \f@ntkey=5 }\textfont\slfam=\eightsl
    \def\bf{\fam\bffam \eightbf \f@ntkey=6 }\textfont\bffam=\eightbf
    \scriptfont\bffam=\eightbf     \scriptscriptfont\bffam=\fivebf
    \def\tt{\fam\ttfam \eighttt \f@ntkey=7 }\textfont\ttfam=\eighttt
    \setbox\strutbox=\hbox{\vrule height 4pt depth 3pt width\z@}
    \samef@nt}
\lock

\def\boxit#1{\vbox{\hrule\hbox{\vrule\kern3pt
             \vbox{\kern3pt#1\kern3pt}\kern3pt\vrule}\hrule}}

\newdimen\str

\def\fboxit#1#2{\vbox{\hrule height #1 \hbox{\vrule width #1
           \kern3pt \vbox{\kern3pt#2\kern3pt}\kern3pt \vrule width #1 }
           \hrule height #1 }}

\def\fillbox#1{\hbox to #1{\vbox to #1{\vfil}\hfil}}
\def\dotbox#1{\hbox to #1{\vbox to 10pt{\vfil}\hss $\cdots$ \hss}}
\def\ggenbox#1#2{\vbox to 10pt{\vss \hbox to #1{\hss #2  \hss} \vss}}


\catcode`\@=11 
\newtoks\foottokens
\let\labelfont=\Tenpoint	
\def\MakeFromBox{\gl@bal\setbox\FromLabelBox=\vbox{\labelfont
     \ialign{##\hfil\cr \the\sendername \the\FromAddress \crcr }}}
\def\smallsize{\relax
\def\eightpoint{\relax
\textfont0=\eightrm  \scriptfont0=\sixrm
\scriptscriptfont0=\fiverm
\def\rm{\fam0 \eightrm \f@ntkey=0}\relax
\textfont1=\eighti  \scriptfont1=\sixi
\scriptscriptfont1=\fivei
\def\oldstyle{\fam1 \eighti \f@ntkey=1}\relax
\textfont2=\eightsy  \scriptfont2=\sixsy
\scriptscriptfont2=\fivesy
\textfont3=\tenex  \scriptfont3=\tenex
\scriptscriptfont3=\tenex
\def\it{\fam\itfam \eightit \f@ntkey=4 }\textfont\itfam=\eightit
\def\sl{\fam\slfam \eightsl \f@ntkey=5 }\textfont\slfam=\eightsl
\def\bf{\fam\bffam \eightbf \f@ntkey=6 }\textfont\bffam=\eightbf
\scriptfont\bffam=\sixbf   \scriptscriptfont\bffam=\sixbf
\def\tt{\fam\ttfam \eighttt \f@ntkey=7 }
\def\caps{\fam\cpfam \tencp \f@ntkey=8 }\textfont\cpfam=\tencp
\setbox\strutbox=\hbox{\vrule height 7.35pt depth 3.02pt width\z@}
\samef@nt}
\normalbaselineskip = 16.60pt plus 0.166pt minus 0.083pt
\normallineskip = 1.25pt plus 0.08pt minus 0.08pt
\normallineskiplimit = 1.25pt
\normaldisplayskip = 16.60pt plus 4.15pt minus 8.3pt
\normaldispshortskip = 4.98pt plus 3.32pt
\normalparskip = 4.98pt plus 1.67pt minus .83pt
\skipregister = 4.15pt plus 1.67pt minus 1.25pt
\def\Eightpoint{\eightpoint \relax
  \ifsingl@\subspaces@t2:5;\else\subspaces@t3:5;\fi
  \ifdoubl@ \multiply\baselineskip by 5
            \divide\baselineskip by 4\fi }
\parindent=16.67pt
\itemsize=25pt
\thinmuskip=2.5mu
\medmuskip=3.33mu plus 1.67mu minus 3.33mu
\thickmuskip=4.17mu plus 4.17mu
\def\thinspace{\kern .13889em }
\def\negthinspace{\kern-.13889em }
\def\enspace{\kern.416667em }
\def\enskip{\hskip.416667em\relax}
\def\quad{\hskip.83333em\relax}
\def\qquad{\hskip1.66667em\relax}
\def\crr{\cropen{8.3333pt}}
\labelwidth=4.5in
\let\labelfont=\Eightpoint
\let\letterhead=\FLOHEAD	
\def\Vfootnote##1{\insert\footins\bgroup
   \interlinepenalty=\interfootnotelinepenalty \floatingpenalty=20000
   \singl@true\doubl@false\Eightpoint
   \splittopskip=\ht\strutbox \boxmaxdepth=\dp\strutbox
   \leftskip=\footindent \rightskip=\z@skip
   \parindent=0.5\footindent \parfillskip=0pt plus 1fil
   \spaceskip=\z@skip \xspaceskip=\z@skip \footnotespecial
   \Textindent{##1}\footstrut\futurelet\next\fo@t}%
\def\attach##1{\step@ver{\strut^{\mkern 1.6667mu ##1} } }
\def\inserttable ##1##2##3%
    {%
    \tbldef {##1}{##3}\goodbreak%
    \midinsert
	\smallskip
	\hbox{\singlespace \hskip 0.5cm
    		\vtop{\parshape=2 0cm 10.8cm 1.3cm 9.5cm
    		      \noindent{\bf\Table{##1}}.\enspace ##3}
		\hfil}
	##2
	\smallskip
    \endinsert
    }
\def\sure{y}
\def\insertfigure ##1##2##3%
    {%
    \figdef {##1}{##3}\goodbreak%
    \midinsert
	\smallskip
	##2
	\hbox{\singlespace\hskip 0.5cm
    		\vtop{\parshape=2 0cm 10.8cm
    			1.6cm 9.2cm \noindent{\bf\Figure{##1}}.
			\enspace ##3}
		\hfil}
	\smallskip
    \endinsert
    }%
\def\references{\par\penalty-300\vskip\chapterskip\spacecheck\chapterminspace
	\line{\twelverm\hfil REFERENCES\hfil}
	\nobreak\vskip\headskip\penalty 30000
	\reflist{}}
\def\figures{\par\penalty-300\vskip\chapterskip\spacecheck\chapterminspace
	\line{\twelverm\hfil FIGURE CAPTIONS\hfil}
	\nobreak\vskip\headskip\penalty 30000
	\figlist{}}
\def\tables{\par\penalty-300\vskip\chapterskip\spacecheck\chapterminspace
	\line{\twelverm\hfil TABLE CAPTIONS\hfil}
	\nobreak\vskip\headskip\penalty 30000
	\tbllist{}}
\def\PH@SR@V{\doubl@true\baselineskip=20.08pt plus .1667pt minus .0833pt
             \parskip = 2.5pt plus 1.6667pt minus .8333pt }
\def\author##1{\vskip\frontpageskip\titlestyle{\tencp ##1}\nobreak}
\def\address##1{\par\kern 4.16667pt\titlestyle{\tenpoint\it ##1}}
\def\andaddress{\par\kern 4.16667pt \centerline{\sl and} \address}
\def\UFL{\address{Department of Physics\break
      University of Florida, Gainesville, FL 32611}}
\def\abstract{\vskip\frontpageskip\centerline{\twelverm ABSTRACT}
              \vskip\headskip }
\def\submit##1{\par\nobreak\vfil\nobreak\medskip
   \centerline{Submitted to \sl ##1}}
\def\doeack{\foot{Work supported by the Department of Energy,
      	contract  DE--FG05--86ER--40272.}}
\def\nsfack{\foot{Work supported by National Science Foundation
      	Grant  PHY 84--16030A01.}}
\def\cases##1{\left\{\,\vcenter{\Tenpoint\m@th
    \ialign{$####\hfil$&\quad####\hfil\crcr##1\crcr}}\right.}
\def\matrix##1{\,\vcenter{\Tenpoint\m@th
    \ialign{\hfil$####$\hfil&&\quad\hfil$####$\hfil\crcr
      \mathstrut\crcr\noalign{\kern-\baselineskip}
     ##1\crcr\mathstrut\crcr\noalign{\kern-\baselineskip}}}\,}
\Tenpoint
}
\newdimen\fullhsize
\newbox\leftcolumn
\def\twoinone{   
\smallsize
\def\papersize{
		\voffset=-.23truein
		\vsize=7truein
		\baselineskip=16pt plus 2pt minus 1pt
		\fullhsize=10truein\hsize=4.75truein\hoffset=-.54truein
		\skip\footins=\bigskipamount}
\def\lettersize{\voffset=.31truein
	        \vsize=6.38truein
		\baselineskip=16pt plus 2pt minus 1pt
		\fullhsize=10truein\hsize=4.75truein\hoffset=-.48truein
		\skip\footins=\smallskipamount \multiply\skip\footins by3}
\papers		
\let\lr=L
\output={\if L\lr
		\global\setbox\leftcolumn=\columnbox \advancepageno
		\global\let\lr=R
	 \else  \getitout \advancepageno
		\global\let\lr=L\fi
	 \ifnum\outputpenalty>-20000 \else\dosupereject\fi}
}		
	\def\columnbox{\leftline{
		\vbox{\ifletterstyle\makeheadline\fi
			\pagebody\makefootline}}}
	\def\fullline{\hbox to\fullhsize}
	\def\getitout{\shipout\vbox{\fullline{\box\leftcolumn
		\hfil {\leftline{
		\vbox{\makeheadline
		\pagebody\makefootline}}} }}}
%
\catcode`\@=12 
%
%
%
\newcount	 \ObjClass
\chardef\ClassNum	= 0
\chardef\ClassMisc	= 1
\chardef\ClassEqn	= 2
\chardef\ClassRef	= 3
\chardef\ClassFig	= 4
\chardef\ClassTbl	= 5
\chardef\ClassThm	= 6
\chardef\ClassStyle     = 7
\chardef\ClassDef       = 8
\edef\NumObj	{\ObjClass = \ClassNum   \relax}
\edef\MiscObj	{\ObjClass = \ClassMisc  \relax}
\edef\EqnObj	{\ObjClass = \ClassEqn   \relax}
\edef\RefObj	{\ObjClass = \ClassRef   \relax}
\edef\FigObj	{\ObjClass = \ClassFig   \relax}
\edef\TblObj	{\ObjClass = \ClassTbl   \relax}

\edef\StyleObj  {\ObjClass = \ClassStyle \relax}
\edef\DefObj    {\ObjClass = \ClassDef   \relax}
%
%
\def\gobble	 #1{}%
\def\trimspace   #1 \end{#1}%
\def\ifundefined #1{\expandafter \ifx \csname#1\endcsname \relax}%
\def\trimprefix  #1_#2\end{\expandafter \string \csname #2\endcsname}%
\def\skipspace #1#2#3\end%
    {%
    \def \temp {#2}%
    \ifx \temp\space \skipspace #1#3\end
    \else \gdef #1{#2#3}\fi
    }%
\def\stylename#1{\expandafter\expandafter\expandafter
    \gobble\expandafter\string\the#1}
\ifundefined {protect} \let\protect=\relax \fi
\catcode`\@=11
\let\rel@x=\relax
\def\relaxtest{\rel@x}
\catcode`\@=12
\def\checkchapterlabel%
    {%
        {\protect\if\chapterlabel\relaxtest
	\global\let\chapterlabel=\relax\fi}
    }%
\begingroup
\catcode`\<=1 \catcode`\{=12
\catcode`\>=2 \catcode`\}=12
\xdef\LBrace<{>%
\xdef\RBrace<}>%
\endgroup
%
%
\newcount\equanumber \equanumber=0
\newcount\eqnumber   \eqnumber=0
\newif\ifleftnumbers \leftnumbersfalse

\def\(#1)%
     {%
        \ifnum \equanumber<0 \eqnumber=-\equanumber
	    \advance\eqnumber by -1 \else
            \eqnumber = \equanumber\fi
        \ifmmode\ifinner(\eqnum{#1})\else
        \ifleftnumbers\leqno(\eqnum{#1})\ifdraft{\rm[#1]}\fi
            \else\eqno(\eqnum{#1})\ifdraft{\rm[#1]}\fi\fi\fi
	\else(\eqnum{#1})\fi\ifnum%
	    \equanumber<0 \global\equanumber=-\eqnumber\global\advance
            \equanumber by -1\else\global\equanumber=\eqnumber\fi
     }%
\def\mideq(#1)%
     {%
	\ifleftnumbers \leqinsert{$\(#1)$} \else
	\eqinsert{$\(#1)$} \fi
     }%
\def\eqnum #1%
    {%
    \LookUp Eq_#1 \using\eqnumber\neweqnum
    {\rm\label}%
    }%
\def\neweqnum #1#2%
    {%
    \checkchapterlabel
    {\protect\xdef\eqnoprefix{\ifundefined{chapterlabel}
	\else\chapterlabel.\fi}}
    \ifmmode \xdef #1{\eqnoprefix #1}
        \else\message{Undefined equation \string#1 in non-math mode.}%
	     \xdef #1{\relax}
	     \global\advance \eqnumber by -1
        \fi
    \EqnObj \SaveObject{#1}{#2}
    }%
\everydisplay = {\expandafter \let\csname Eq_\endcsname=\relax
		 \expandafter \let\csname Eq_?\endcsname=\relax}%
%
%
\newcount\tablecount \tablecount=0
\def\Table  #1{Table~\tblnum {#1}}%
\def\tblnum #1{\TblObj \LookUp Tbl_#1 \using\tablecount
	\SaveObject \label\ifdraft [#1]\fi}%
\def\tbldef #1{\TblObj \SaveContents {Tbl_#1}}%
\def\tbllist  {\TblObj \ListObjects}%
%
\def\inserttable #1#2#3%
    {%
    \tbldef {#1}{#3}\goodbreak%
    \midinsert
	\smallskip
	\hbox{\singlespace
	      \vtop{\titlestyle{{\Tenpoint{\caps\Table{#1}}\break #3}}}
    	     }%
	#2
    	\smallskip%
    \endinsert
    }%
\def\topinserttable #1#2#3%
    {%
    \tbldef {#1}{#3}\goodbreak%
    \topinsert
	\smallskip
	\hbox{\singlespace
	      \vtop{\titlestyle{{\Tenpoint{\caps\Table{#1}}\break #3}}}
    	     }%
	#2
    	\smallskip%
    \endinsert
    }%
%
%
\newcount\figurecount \figurecount=0
\def\Figure #1{Figure~\fignum {#1}}%
\def\Fig    #1{Fig.~\fignum {#1}}%
\def\fignum #1{\FigObj \LookUp Fig_#1 \using\figurecount
     \SaveObject \label\ifdraft [#1]\fi}%
\def\figdef #1{\FigObj \SaveContents {Fig_#1}}%
\def\figlist  {\FigObj \ListObjects}%
%
\def\sure{y}
\def\insertfigure #1#2#3%
    {%
    \figdef {#1}{#3}%
    \midinsert
	\bigskip
	\ifx\havefigures\sure
	#2
	\else\fi
	\hbox{	\singlespace
		\hskip 0.4in
		\vtop{\parshape=2 0pt 362pt 32pt 330pt
		      \noindent{\Tenpoint{\caps\Fig{#1}}.\enspace #3}}
		\hfil}
    	\smallskip%
    \endinsert
    }%
\def\topinsertfigure #1#2#3%
    {%
    \figdef {#1}{#3}%
    \topinsert
	\bigskip
	#2
	\hbox{	\singlespace
		\hskip 0.4in
		\vtop{\parshape=2 0pt 362pt 32pt 330pt
		      \noindent{\Tenpoint{\caps\Fig{#1}}.\enspace #3}}
		\hfil}
    	\smallskip%
    \endinsert
    }%
%
%
%
\newcount\theoremcount \theoremcount=0
%
%
%
%
%
%
%
%
%
%
%
%
%
%
\newcount\referencecount \referencecount=0
\newcount\refsequence	\refsequence=0
\newcount\lastrefno	\lastrefno=-1
%
\def\NPrefs{\let\refmark=\NPrefmark \let\refitem=\NPrefitem}

%
\def\refsymbol#1{\refrange#1-\end}%
\def\[#1]#2%
	{%
	\if.#2\rlap.\refmark{\refsymbol{#1}}\let\refendtok=\relax%
	\else\if,#2\rlap,\refmark{\refsymbol{#1}}\let\refendtok=\relax%
    	\else\refmark{\refsymbol{#1}}\let\refendtok=#2\fi\fi%
	\discretionary{}{}{}\refendtok}%
\def\refrange #1-#2\end%
    {%
    \refnums #1,\end
    \def \temp {#2}%
    \ifx \temp\empty \else -\expandafter\refrange \temp\end \fi
    }%
\def\refnums #1,#2\end%
    {%
    \def \temp {#1}%
    \ifx \temp\empty \else \skipspace \temp#1\end\fi
    \ifx \temp\empty
	\ifcase \refsequence
	    \or\or ,\number\lastrefno
	    \else  -\number\lastrefno
	\fi
	\global\lastrefno = -1
	\global\refsequence = 0
    \else
	\RefObj \edef\temp {Ref_\temp\space}%
	\expandafter \LookUp \temp \using\referencecount\SaveObject
	\global\advance \lastrefno by 1
	\edef \temp {\number\lastrefno}%
	\ifx \label\temp
	    \global\advance\refsequence by 1
	\else
	    \global\advance\lastrefno by -1
	    \ifcase \refsequence
		\or ,%
		\or ,\number\lastrefno,%
	    \else   -\number\lastrefno,%
	    \fi
	    \label
	    \global\refsequence = 1
	    \ifx\suffix\empty
		\global\lastrefno = \label
	    \else
		\global\lastrefno = -1
	    \fi
	\fi
	\refnums #2,\end
    \fi
    }%
%
%
%
%
\def\refdef #1{\RefObj \SaveContents {Ref_#1}}%
\def\reflist  {\RefObj \ListObjects}%
%
%
%
%
\newif\ifSaveFile
\newif\ifnotskip
\newwrite\SaveFile
\let\IfSelect=\iftrue
\edef\savefilename {\jobname.aux}%
\def\Def#1#2%
    {%
    \expandafter\gdef\noexpand#1{#2}%
    \DefObj \SaveObject {#2}{\expandafter\gobble\string#1}%
}%
\def\savestate%
    {%
    \ifundefined {chapternumber} \else
	\NumObj \SaveObject {\number\chapternumber}{chapternumber} \fi
        \ifundefined {appendixnumber} \else
	\NumObj \SaveObject {\number\appendixnumber}{appendixnumber} \fi
    \ifundefined {sectionnumber} \else
	\NumObj \SaveObject {\number\sectionnumber}{sectionnumber} \fi
    \ifundefined {pagenumber} \else
	\advance\pagenumber by 1
	\NumObj \SaveObject {\number\pagenumber}{pagenumber}%
	\advance\pagenumber by -1 \fi
    \NumObj \SaveObject {\number\equanumber}{equanumber}%
    \NumObj \SaveObject {\number\tablecount}{tablecount}%
    \NumObj \SaveObject {\number\figurecount}{figurecount}%
    \NumObj \SaveObject {\number\theoremcount}{theoremcount}%
    \NumObj \SaveObject {\number\referencecount}{referencecount}%
    \checkchapterlabel
    \ifundefined {chapterlabel} \else
	{\protect\xdef\chaplabel{\chapterlabel}}
	\MiscObj \SaveObject \chaplabel {chapterlabel} \fi
    \ifundefined {chapterstyle} \else
   	\StyleObj \SaveObject {\stylename{\chapterstyle}}{chapterstyle} \fi
    \ifundefined {appendixstyle} \else
	\StyleObj \SaveObject {\stylename{\appendixstyle}}{appendixstyle}\fi
}%
\def\Contents #1{\ObjClass=-#1 \SaveContents}%
\def\Define #1#2#3%
    {%
    \ifnum #1=\ClassNum
	\global \csname#2\endcsname = #3 %
    \else \ifnum #1=\ClassStyle
	\global \csname#2\endcsname\expandafter=
	\expandafter{\csname#3\endcsname} %
    \else \ifnum #1=\ClassDef
        \expandafter\gdef\csname#2\endcsname{#3} %
    \else
	\expandafter\xdef \csname#2\endcsname {#3} \fi\fi\fi %
    \ObjClass=#1 \SaveObject {#3}{#2}%
    }%
\def\SaveObject #1#2%
    {%
    \ifSaveFile \else \OpenSaveFile \fi
    \immediate\write\SaveFile
	{%
	\noexpand\IfSelect\noexpand\Define
	{\the\ObjClass}{#2}{#1}\noexpand\fi
	}%
    }%
\def\SaveContents #1%
    {%
    \ifSaveFile \else \OpenSaveFile \fi
    \BreakLine
    \SaveLine {#1}%
    }%
\begingroup
    \catcode`\^^M=\active %
\gdef\BreakLine %
    {%
    \begingroup %
    \catcode`\^^M=\active %
    \newlinechar=`\^^M %
    }%
\gdef\SaveLine #1#2%
    {%
    \toks255={#2}%
    \immediate\write\SaveFile %
	{%
	\noexpand\IfSelect\noexpand\Contents
	{-\the\ObjClass}{#1}\LBrace\the\toks255\RBrace\noexpand\fi%
	}%
    \endgroup %
    }%
\endgroup
\def\ListObjects #1%
    {%
    \ifSaveFile \CloseSaveFile \fi
    \let \IfSelect=\GetContents \ReadFileList #1,\savefilename,\end
    \let \IfSelect=\IfDoObject  \input \savefilename
    \let \IfSelect=\iftrue
    }%
\def\ReadFileList #1,#2\end%
    {%
    \def \temp {#1}%
    \ifx \temp\empty \else \skipspace \temp#1\end \fi
    \ifx \temp\empty \else \input #1 \fi
    \def \temp {#2}%
    \ifx \temp\empty \else \ReadFileList #2\end \fi
    }%
\def\GetContents #1#2#3%
    {%
    \notskipfalse
    \ifnum \ObjClass=-#2
	\expandafter\ifx \csname #3\endcsname \relax \else \notskiptrue \fi
    \fi
    \ifnotskip \expandafter \DefContents \csname #3_\endcsname
    }%
\def\DefContents #1#2{\toks255={#2} \xdef #1{\the\toks255}}%
\def\IfDoObject #1#2%
    {%
    \notskipfalse \ifnum \ObjClass=#2 \notskiptrue\fi \ifnotskip \DoObject
    }%
\def\DoObject #1#2%
    {%
    \ifnum \ObjClass = \ClassTbl	\par\noindent Table~#2.
    \else \ifnum \ObjClass = \ClassFig	\par\noindent Figure~#2.
    \else \ifnum \ObjClass = \ClassRef  \refitem{#2}
    \else \item {#2.}
    \fi\fi\fi
    \ifdraft\edef\temp {\trimprefix #1\end}[\expandafter\gobble \temp]~\fi
    \expandafter\ifx \csname #1_\endcsname \relax
	\ifdraft\relax\else\edef\temp {\trimprefix #1\end}%
	[\expandafter\gobble \temp]\fi%
    \else
	\csname #1_\endcsname
    \fi
    }%
\def\OpenSaveFile   {\immediate\openout\SaveFile=\savefilename
		     \global\SaveFiletrue}%
\def\CloseSaveFile  {\immediate\closeout\SaveFile \global\SaveFilefalse}%
%
%
\def\LookUp #1 #2\using#3#4%
    {%
    \expandafter \ifx\csname#1\endcsname \relax
	\global\advance #3 by 1
	\expandafter \xdef \csname#1\endcsname {\number #3}%
	\let \newlabelfcn=#4%
	\ifx \newlabelfcn\relax \else
	    \expandafter \newlabelfcn \csname#1\endcsname {#1}%
	\fi
    \fi
    \xdef \label  {\csname#1\endcsname}%
    \gdef \suffix {#2}%
    \ifx \suffix\empty \else
	\xdef \suffix {\expandafter\trimspace \suffix\end}%
	\xdef \label  {\label\suffix}%
    \fi
    }%
%
%
%
\newcount\appendixnumber	\appendixnumber=0
\newtoks\appendixstyle		\appendixstyle={\Alphabetic}
\newif\ifappendixlabel		\appendixlabelfalse
\def\APPEND#1{\par\penalty-300\vskip\chapterskip\spacecheck\chapterminspace
        \global\chapternumber=\number\appendixnumber
	\global\advance\appendixnumber by 1
	\chapterstyle\expandafter=\expandafter{\the\appendixstyle} \chapterreset
	\titlestyle{Appendix\ifappendixlabel~\chapterlabel\fi.~ {#1}}
	\nobreak\vskip\headskip\penalty 30000}
%

%
%
%
\def\references#1{\par\penalty-300\vskip\chapterskip\spacecheck
	\chapterminspace\line{\fourteenrm\hfil References\hfil}
	\nobreak\vskip\headskip\penalty 30000\reflist{#1}}
\def\figures#1{\par\penalty-300\vskip\chapterskip\spacecheck
	\chapterminspace\line{\fourteenrm\hfil Figure Captions\hfil}
	\nobreak\vskip\headskip\penalty 30000\figlist{#1}}
\def\tables#1{\par\penalty-300\vskip\chapterskip\spacecheck
	\chapterminspace\line{\fourteenrm\hfil Table Captions\hfil}
	\nobreak\vskip\headskip\penalty 30000\tbllist{#1}}
\newif\ifdraft\draftfalse
\newcount\yearltd\yearltd=\year\advance\yearltd by -1900
\def\draft{\drafttrue
	\def\draftdate{preliminary draft:
		\number\month/\number\day/\number\yearltd\ \ \hourmin}%
	\paperheadline={\hfil\draftdate} \headline=\paperheadline
	{\count255=\time\divide\count255 by 60 \xdef\hourmin{\number\count255}
        	\multiply\count255 by-60\advance\count255 by\time
		\xdef\hourmin{\hourmin:\ifnum\count255<10 0\fi\the\count255} }
	\message{draft mode}  }
%

\def\umdepp{ \Pubnum={$\caps UMD - EPP - \the\pubnum $}
	     \pubtype={$\phantom{blah}$}                 }
\def\slacpub{ \Pubnum={$\caps SLAC - PUB - \the\pubnum $}}
%
%

\def\frac#1/#2{\leavevmode\kern.1em\raise.5ex
		\hbox{\the\scriptfont0
         	#1}\kern-.1em/\kern-.15em
		\lower.25ex\hbox{\the\scriptfont0 #2}}
%
%

\def\frac#1#2{{#1 \over #2}}
\def\goesas#1#2{\ \ {\phantom{a} \atop
                \widetilde{#1\rightarrow #2}}}
\def\half{{\frac 12}}
\def\third{{\frac 13}}

\def\eq#1{$\(#1)$}
 \refdef{hoyertw}{P. Hoyer, N. A. T\"ornqvist and B. R. Webber, Phys. Lett.
61B (1976) 191; Nucl. Phys. B115 (1976) 429.}
\refdef{hoyer}{P. Hoyer, Phys. Lett. 63B (1976) 50.}
\refdef{cerulusm} {F. Cerulus and A. Martin,
{\sl Phys. Lett.} {\bf 8} (1964) 80.}
\refdef{grossmende} {D. J. Gross and P. F. Mende,
{\sl Phys. Lett.} {\bf 197B} (1987) 129;
{\sl Nucl. Phys.} {\bf B303} (1988) 407.}
\refdef{curtrightgt}{T. L. Curtright, C. B. Thorn, and J. Goldstone,
{\sl Phys. Lett.} {\bf 175B} (1986) 47;
T. L. Curtright and C. B. Thorn,
{\sl Nucl. Phys.} {\bf B274} (1986) 520.}
\Pubnum = {UFIFT-94-2\cr
}
\date = {}

\titlepage
\def\doeack{\foot{Work supported by the Department of Energy,
      	contract  DE--FG05--86ER--40272.}}
\title{Interpretation of High Energy String
Scattering in terms of String Configurations\doeack}
\author{Steven L. Carbon and Charles B. Thorn}
\address{Department of Physics\break
University of Florida, Gainesville, FL 32611}
\abstract
High energy string scattering at fixed momentum transfer,
known to be dominated by Regge trajectory exchange, is interpreted
by identifying families of string states which induce each type
of trajectory exchange. These include the usual leading trajectory
$\alpha(t)=\alpha^\prime t+1$ and its daughters
as well as the ``sister'' trajectories
$\alpha_m(t)=\alpha(t)/m-(m-1)/2$ and their daughters. The
contribution of the sister $\alpha_m$ to high energy scattering
is dominated by string excitations
in the $m^{th}$ mode. Thus, at large
$-t$, string scattering is dominated by wee partons, consistently
with a picture of string as an infinitely composite system of
``constituents'' which carry zero energy and momentum.
\pagenumbers
\endpage

Superstring theory is apparently not an
ordinary quantum field theory, and
yet it seems to satisfy all of the axioms of $S$-matrix theory.
It is also the only existing physical theory of quantum gravity
which has a chance
of avoiding the notorious ultraviolet divergences of quantized
Einstein gravity,
a disease which also afflicts
all of the supersymmetric extensions of
that theory. Unfortunately, we do not yet know how to
properly formulate string theory. String field theory is an
option  that is still being pursued, but it must be said that
the concept of a string field is, to say the least,  cumbersome.
All of perturbation theory can be so simply understood using the
first-quantized world sheet approach, that one feels
there must be a better way.

The string field concept is
derived directly from the perturbative
particle content of string theory: one assigns an independent
string field component to each state of the string. It has long
been recognized that the field content of quantum field theory
has no necessary correspondence to the particle content. Just
think of QCD where no fundamental field is associated with any
particle state. Why should it be any different for string theory?
If we knew only the particle states (\ie\ the hadrons) of QCD,
we could apply a familiar strategy for uncovering the fundamental
degrees of freedom: probing the short distance structure of hadrons
reveals the fundamental degrees of freedom, namely
quarks and gluons. In this article we apply the
same strategy to string theory. The string scattering amplitudes
are known in perturbation theory.
We shall try to use them to illuminate
the short distance structure of string theory.

To probe short distances, it is necessary to study scattering
amplitudes at very large momentum. It would be desirable to
consider taking all components of all momenta large, but this
is impossible for $S$-matrix elements since the external lines
are all on mass-shell. For example, consider a four particle
scattering amplitude. There are only two independent invariants,
$s=-(p_1+p_2)^2$ and $t=-(p_4+p_1)^2$. All others are related
to these via momentum conservation and the mass-shell conditions,
\eg\ $u=-(p_1+p_3)^2=\Sigma-s-t$
where $\Sigma=\sum_{k=1}^4m_k^2$.
Gross and Mende\[grossmende] have
already studied the maximally high momentum
limit where all three
invariants $s, t, u$ become large in magnitude.
This is achieved by taking high energy at fixed scattering angle.
The striking conclusion is that the amplitudes decay exponentially
in all these invariants, in sharp contrast to the power law
fall off of all known quantum field theories. If this exponential
decay is associated with the vertex form factors, the result
could be interpreted as the scattering of two softly bound
composite systems. The result, by itself, does not necessarily
imply that any hypothetical constituents of the composites are
themselves soft. This latter conclusion is expected from the
naive form factor estimate for a charge attached to a point
on the string
$$F(Q^2)\sim e^{-\alpha^\prime Q^2\ln n_C}$$
and the electroproduction structure function
$$W(x_{Bj})\sim e^{-x_{Bj}n_C},$$
where we have arbitrarily suppressed the
contribution of string oscillation modes $n>n_C$.
The coefficient of $x_{Bj}$ in the exponent is infinite
when the cutoff is
removed ($n_C\rightarrow\infty$), indicating zero probability
for finding the point charge with finite momentum.
(Recall that $x_{Bj}$ equals the momentum fraction carried by the
struck constituent.) Since we
really don't know how local fields should couple to a
string, one of our aims in this article is to expose a
true $S$-matrix signature of this wee parton dominance.

Instead of the extreme short
distance limit studied by Gross and Mende,
in this article we consider large momentum limits in
which a maximal number of invariants are large, but
some are much larger than
others. Such limits probe the Regge trajectories of string
theory and hence might shed light on the more stringy
aspects of the short distance limit. Consider the contribution
of the leading Regge trajectory to the four open string scattering
amplitude (with no Chan-Paton factors):
$$A_4(s,t)\goesas s\infty {1+e^{-i\pi\alpha(t)}\over2}
\Gamma(-\alpha(t))s^{\alpha(t)}.$$
where
$$\alpha(t)=\alpha^\prime t+1.$$
As $t$ varies from 0 to $-\infty$ these Regge contributions
interpolate between the stringy particle spectrum and the
high energy fixed angle limit.

Long ago Hoyer, Tornquist and Webber\[hoyertw,hoyer] discovered
that in addition to the familiar linear Regge trajectories, there
is an infinite sequence of sister trajectories (see \Fig{sisters})
$$\alpha_n(t)={1\over n}\alpha(t)-{n-1\over2}$$
with ever decreasing slope which contribute to
high energy string processes. These new trajectories couple only to
higher point functions (typically $\alpha_n$ contributes
only to amplitudes involving at least $2(n+1)$ strings),
but their existence nonetheless suggests an interesting
modification of the approach to fixed angle scattering
described above. In particular, although each individual
trajectory decreases linearly with $t$ as $t\rightarrow-\infty$,
the envelope of the complete family of trajectories decreases
only as $-\sqrt{-t}$ in the same limit.
If the Cerulus-Martin lower
bound\[cerulusm] on fixed angle scattering cross sections
$$|A(s,t)|>e^{-f(\theta)\sqrt{s}\ln s}$$
holds, it follows that the scattering amplitude can not be
dominated by a Regge trajectory $\alpha(t)$ for {\it all}
$t<0$ unless that trajectory satisfies $\alpha(t)>-C\sqrt{-t}$
with $C>0$ as $t\rightarrow-\infty$. Thus if the envelope of
sisters is taken as the effective leading trajectory for the
tree approximation to string theory, we can say that the tree
approximation is in this sense compatible with the bound.
In multi-loop string amplitudes an effect similar to that of
the sister trajectories is produced by multi-Regge cuts, so
in their work,
Gross and Mende anticipate that multiloop corrections could
restore the bound. We are raising here the possibility that
the ``classical'' string theory (\ie\ tree approximation)
contains a mechanism for restoring
the bound, \eg\ by exploiting a different
background.

\insertfigure{sisters}{
\centerline{\psfig{figure=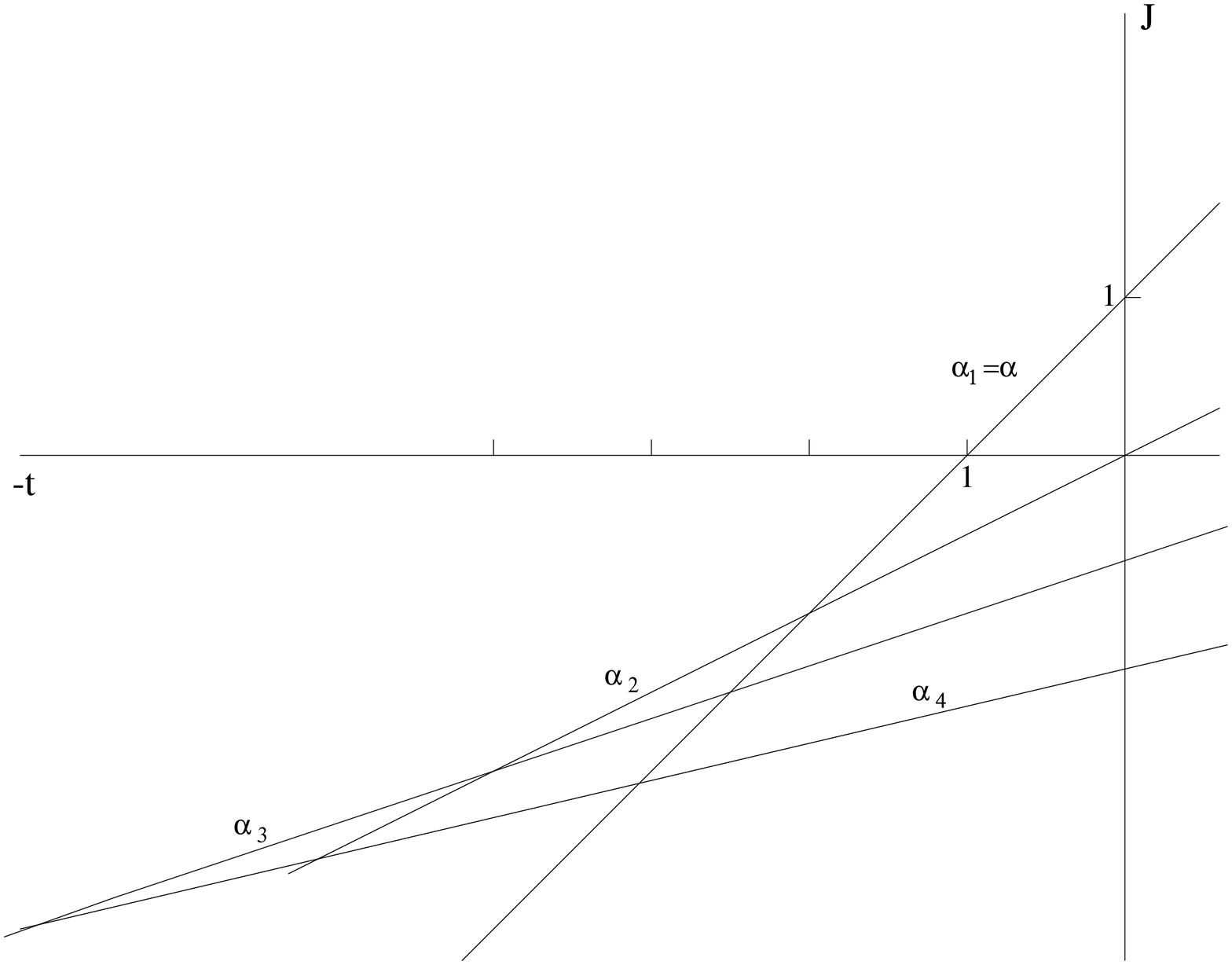,height=8cm}}}
{The first four leading sister trajectories. Units are chosen so
$\alpha^\prime=1$.}

There are two approaches toward extracting Regge trajectories from
scattering amplitudes. The first is to analyse directly the high
energy limit at fixed $t$ and to extract the trajectory as the
power of $s$ that characterizes the asymptotics. This is the
only way the sisters have been obtained so far. Unfortunately
this analysis obscures the string interpretation. The second method
is to look for angular momentum singularities in the partial
wave scattering amplitudes. Since these angular momentum
singularities are directly associated with the excitation spectrum
of the string, the string interpretation should be more transparent
in this second approach.
In this note we shall explain the origin of the sister trajectories
in terms of the sum over string excited states contributing as
intermediate resonances in the scattering amplitudes. This analysis
allows a more direct interpretation of them in terms of string
configurations. It also provides a link with an earlier analysis
of the spin content of the physical states of string
theory\[curtrightgt] in which this
pattern of sister trajectories naturally appears. Our work
confirms the relevance of the spin content analysis to the
high energy behavior of string scattering amplitudes.
Since the whole family of sister trajectories
encodes information about the short distance structure of string
theory, we hope that this interpretation will yield hints about
the relevant degrees of freedom that might be used in a reformulation
of string theory.

Instead of the usual procedure of
trying to define explicit partial wave
projections on multiparticle scattering
amplitudes, we can directly
exploit the string oscillator formalism to display the poles
in any selected channel as a sum over string states according to
the decomposition of the identity
$$I= \sum_{n_1=0}^\infty {1\over n_1!}
\sum_{n_2=0}^\infty {1\over 2n_2!}
  \cdots \biggl(a_{-1}^{n_1} a_{-2}^{n_2} \cdots \mid 0\rangle
  \cdot\langle 0\mid \cdots a_2^{n_2} a_1^{n_1}\biggr).\(id)$$
The normalizations in \eq{id} are
fixed by the projector condition $I^2=I$,
and the commutation relations
$$[a_m^\mu,a_n^\nu] = m \delta_{m+n}\eta^{\mu\nu}.\()$$
We have also suppressed the $\mu_k$ indices: all indices in the ket
are understood to be contracted with corresponding indices in the bra.
Pick a channel in $M$ open string scattering with total momentum $P$
 and consider the dependence on the invariant
$t=-P^2$. The poles in the scattering amplitude can then be
exhibited as
$$A_M =\langle L\mid\prod_{r=1}^\infty \sum_{n_r=0}^\infty
  {1\over r^{n_r} n_r!}
a_{-r}^{n_r}\mid 0\rangle\cdot
\langle 0\mid a_r^{n_r}
{1\over {\sum_{j=1}^\infty jn_j-\alpha(t)}}\mid R\rangle.  \(amps1)$$
Our aim is to identify families of states which can be associated
with Regge trajectories of a definite type and then to examine how
each such family generates singularities in the complex
 angular momentum plane.

Let us begin by asking which family of states generates the leading
Regge trajectory $\alpha(t)=\alpha^\prime t+1$. These should be
the states that maximize the angular momentum at fixed
rest mass. For the classical string these would be the motions
in which a straight string rotates about its center
in a plane. For the first quantized string, we
fix a (mass)$^2$ level and notice that we can maximize the angular
momentum by maximizing the number of
oscillator raising operators, which
we do by using only the $a_{-1}^\mu$
operators in constructing the state:
$$a_{-1}^{\mu_1}a_{-1}^{\mu_2}\cdots a_{-1}^{\mu_N}\ket{0,P}
\(family1)$$
where $-\alpha^\prime P^2=N-1$.
In the string's rest system the angular
momentum of these states can be identified by the fact that the
space components of each oscillator transform as an $O(D-1)$
vector and the time component as a scalar. The spatial components
always enter as a completely symmetric $O(D-1)$ tensor, so the
irreducible angular momentum states are formed by removing all
the traces. The maximal angular momentum is clearly $J=N$ and
corresponds to the states with all $\mu_k$ spatial and projected
to be traceless in all spatial indices. We must associate the
leading trajectory with the entire family of these maximal angular
momentum states with $N=0,1,2,\cdots$. Clearly for this family,
$J=N=\alpha^\prime t+1$.

The next step is to examine how this family of states contributes
to the scattering amplitude. To end up with only the leading
trajectory we would have to carry out all of the restrictions
on the $\mu_k$ and the trace projections described in the preceding
paragraph. This would be tedious and, for our purposes, not very
illuminating. If we take the whole family $\(family1)$ instead,
it is clear that we should generate the leading trajectory
accompanied by some lower lying daughter trajectories parallel to the
leading one and spaced by integer values in
angular momentum.
In the relativistic context it is indeed more natural
to consider such groups of trajectories. So let us restrict the
sum over states in \eq{amps1} to \eq{family1}.
$$A_M^{(1)} =\sum_{n=0}^\infty
  {1\over n!}
\langle L\mid a_{-1}^{n}\mid 0\rangle\cdot
\langle 0\mid a_1^{n}
\mid R\rangle {1\over {n-\alpha(t)}}.  \(A1)$$
The sum over $n$ in this formula is closely related to a
partial wave expansion. Instead of being a pure angular momentum,
$n$ is a label denoting a group of angular momentum values
whose maximal value is $n$. But for purposes of interpreting
high energy behavior it is just as effective as the
pure angular momentum. Thus instead of the conventional
Sommerfeld-Watson transformation on a true partial wave expansion,
we will apply the same transformation to (a slightly
rearranged version of) the sum over $n$ in \eq{A1}.
To do this one has to continue the summand to continuous complex
values of $n$. The explicit $n$ dependence shown in \eq{A1} has
a clear continuation. Moreover the ``Regge'' pole
$n=\alpha(t)$ is quite explicit. However the continuation of
the $n$ dependence in the product of matrix elements is not
obvious so we now turn to this question.

Call the momenta of the $I<M$ strings,
described by the state $\ket{R}$,
$p_i$ and those $J=M-I$,
described by the state $\bra{L}$, $q_j$. Each
vertex factor is in the form of a Koba-Nielsen integral over
variables which we call $y_i$ for the right factor and $x_j$
for the left factor. We can choose, for example, $y_1=x_1=0$
and $y_I=x_J=1$ with all other variables taking values between
0 and 1. Define $P_n(y)=\sum_i p_iy_i^n$ and $Q_n=\sum_j q_jx_j^n$.
Then the products of matrix elements in the $n^{th}$ term of
\eq{A1} can be written
$$F_n\equiv\langle L\mid a_{-1}^{n}\mid 0\rangle\cdot
\langle 0\mid a_1^{n}
\mid R\rangle=\int\prod_{i=2}^{I-1}dy_i\prod_{j=2}^{J-1}dx_j
L(x,q)R(y,p)[Q_1(x)\cdot P_1(y)]^n.  \(pdotq)$$
Clearly the $n$ dependence is isolated in the last factor.
This formula shows that for fixed integer $n$ the invariants
$p_i\cdot q_j$ enter through a homogeneous polynomial of order $n$.
These invariants are the multi-particle analogue of the crossed
channel variable $s$ in 2-2 scattering. They are linear in the
cosines of the various scattering angles, and this explains why the
sum over $n$ is analogous to the partial wave expansion. It is
the $p_i\cdot q_j$ which are taken large in Regge limits. For $M>4$
there is some degree of choice in specifying a high energy limit,
and an associated choice of the suitable summation variable to
continue into the complex plane. A simple limit would be to
take all these invariants to infinity at fixed ratio, and if this
were possible the variable $n$ of \eq{A1} is the suitable one.
However, for
a fixed dimension of space-time, this limit is impossible except
for low values of $M$.
In this paper, we consider high energy limits which
are possible in 4 space-time dimensions. That is, if our theory
is defined
in a higher number of dimensions, we shall only allow large momentum
components in a four dimensional subspace. If we let $s$ be
one of the large invariants,
then because of the 4-dimensionality constraints
on the large momentum components entering the $q_i\cdot p_j$,
$Q\cdot P$ will not be homogeneous in $s$ but we can write
$Q\cdot P=As+B$ with $A$ independent of $s$ and $B=O(1)$ as
$s\rightarrow\infty$. Then
$$[Q_1(x)\cdot P_1(y)]^n=\sum_{k=0}^n {n\choose k} (As)^{n-k}B^k,$$
And a rearrangement of the sum in \eq{A1} yields
$$
A^{(1)}_M=\sum_{n=0}^\infty{1\over n!}s^n\sum_{k=0}^\infty{1\over k!}
{1\over n+k-\alpha(t)}F^k_n \(refined)$$
where $F^k_n$ is obtained from $F_n$ by replacing $[Q_1\cdot P_1]^n$
by $A^nB^k$ in \eq{pdotq}. We stress that the choice of $B$ is
determined by the high energy limit we wish to analyze. Although
the limit determining the choice $B=0$ might be feasible for
low $M$, it would only be possible {\it for all
$M$} in an infinite number of space-time dimensions. Note also that
the term containing the leading Regge pole is the one with $k=0$.

Turning at last to the problem of continuing $n$, we first note
that for integer $n$, $A$ is just a polynomial in
$y$ and $x$ and has a benign effect on the integrals; in
particular the vanishing of $A$ in the integration
range causes no difficulty. However when we attempt to
continue $n$ to continuous complex values, the vanishing
of $A$ in the integration range will induce singularities in
the $n$ dependence at negative values of $n$. Before discussing
these singularities we must first spell out how the continuation
in $n$ is to be done. Carlson's theorem
shows that such a continuation is unique up to ambiguities
like $\sin n\pi$. If $A$ changes sign in the integrand,
our expression is plagued by a dependence $(-)^n$ which is
just of this ambiguous type. This difficulty can be handled
by defining separate continuations for even and odd $n$:
$$A^n=\cases{|A|^n & $n$ even\cr
A|A|^{n-1} & $n$ odd.\cr}$$
This is completely analogous to the usual definition of
signatured amplitudes in Regge theory.

Call the continuation
of $F^k_n$ for even $n$, $F^{k(+)}(n)$,
and that for odd $n$, $F^{k(-)}(n)$. Then the Sommerfeld-Watson
representation for the r.h.s. of \eq{refined} reads
$$A_M^{(1)} ={1\over2\pi i}\int_{C_0} dz
  {s^z\over \sin\pi z\Gamma(z+1)}\sum_{k=0}^\infty{1\over k!}
{\xi_+(z)F^{k(+)}(z)+\xi_-(z)F^{k(-)}(z)\over {z+k-\alpha(t)}}
  \(SW1)$$
Here the contour $C_0$ is a set of infinitesimal circles around
all the nonnegative integers. The signature factors
$\xi_{\pm}={(e^{-i\pi z}\pm1)/2}$ guarantee that the signatured
amplitudes contribute only where they should: $F^{(\pm)}$ for
even(odd) $n$ respectively. The Sommerfeld-Watson transformation
consists in deforming the contour $C_0$ as far as possible to
the left in the complex $z$ plane. Singularities exposed by this
deformation then contribute to the large $s$ behavior,
the rightmost one dominating. Inspection of \eq{SW1}
shows that there is always the ``normal'' Regge behavior
coming from picking up the explicit poles at $z=\alpha(t)-k$.
In a situation where $A$ never vanishes (for example,
when all the $p_i\cdot q_j$ have the same sign) there are no
further singularities in $z$ and one has only normal
Regge behavior.

However, generally the $p_i\cdot q_j$ have both signs and
$A$ vanishes inside the integration region. Then integration
near such a zero
induces further singularities that are not of the ``normal''
Regge type. In the truncated amplitude $A^{(1)}$ the new singularities
are fixed (independent of $t$) and can give rise to fixed power
behavior.\foot{Singularities arising from an interior
region of integration typically occur at negative (``nonsense'')
``wrong signature'' integers. Thus the signature
factor in the Sommerfeld-Watson representation supplies a
zero so {\it at least} a double pole singularity is required
to give rise to fixed power behavior.}
Thus they can modify the expected rate at which
the amplitude
vanishes with $s$ at large $t$. The occurrence of these
fixed singularities in $z$
depends on the process (\eg\ for $M=4$ no such singularity occurs),
and on which high energy limit is taken. We
restrict attention to high energy
limits in which only momentum components within a four-dimensional
subspace of space-time
are allowed to become large. The main consequence
of this restriction is that there are relations among
the invariants $p_i\cdot q_j$ which are becoming large, and,
in particular $B\neq0$ for $M\geq6$.

Let us confirm for $M=6$ that a double pole at $z=-1$ appears.
In this case ($I=J=3$, $x_2\equiv x, y_2\equiv y$), putting
$s=p_2\cdot q_2$, $\eta_1=p_3\cdot q_2/s$, and
$\eta_2=p_2\cdot q_3/s$,
$$\eqalign{
P\cdot Q=&p_3\cdot q_3+xp_3\cdot q_2+yp_2\cdot q_3+yxp_2\cdot q_2\cr
\equiv& s\left(y-\eta_1\right)
\left(x-\eta_2\right)
+\left\{{p_3\cdot q_3 p_2\cdot q_2-q_3\cdot p_2 p_3\cdot q_2\over
s}\right\}\cr}
$$
In a 4 dimensional high energy
limit there is a relation between the invariants
that forces $\kappa$,
the quantity in braces, to be $O(1)$ for large $s$ so it
plays the role of $B$ in our earlier discussion. The role of $A$ is
played by the factored
expression $(y-\eta_1)(x-\eta_2)$. Thus we have for
$M=6$
$$F^{k(+)}(z)=\kappa^k\int_0^1 dy R(y,p)|y-\eta_1|^z\int_0^1 dx
L(x,q)|x-\eta_2|^z$$
If $0<\eta_1, \eta_2<1$ we see that integration near $y\approx\eta_1$,
$x\approx\eta_2$ produces a double pole at $z=-1$, a nonsense
wrong-signature point,
$$F^{k(+)}\goesas z {-1}
{4\kappa^k R(\eta_1,p)L(\eta_2,q)\over(z+1)^2},$$
which contributes the high $s$ behavior
$${2i\over s}R(\eta_1,p)L(\eta_2,q)\sum_{k=0}^\infty{\kappa^k\over k!}
{1\over k-1-\alpha(t)}\(fixedpole)$$
to $A_6^{(1)}$. There is no
singularity at $z=-1$ in $F^{k(-)}(z)$.

The contribution \eq{fixedpole}
to the high energy behavior of $A^{(1)}$
is intimately linked to the first sister trajectory $\alpha_2(t)$.
At the moment it appears to be a fixed singularity in the
angular momentum plane, but that is because we have only included
a subset of the intermediate states \eq{family1}. The complete
set of states would be obtained by including all states of the
same form as \eq{family1} but with the ket $\ket{0,P}$ replaced
by any state obtained by applying an arbitrary monomial
of the $a_{-n}$ with $n\neq 1$ to $\ket{0,P}$:
$$a_{-2}^{n_2}a_{-3}^{n_3}
\cdots\ket{0,P}\(comp1),$$
where the integers $n_k$ represent powers of the $k^{th}$ oscillators
and Lorentz indices are suppressed.
Thus the total contribution from the double pole at $z=-1$
will itself be a sum over intermediate states of the form \eq{comp1}.
The pole factor for each such contribution will be
$1/(-1+2n_2+3n_3\cdots-\alpha(t))$ and the leading power of
$s$ will be $s^{-1+n_2+n_3\cdots}$. The states with maximal
angular momentum amongst the states \eq{comp1} are those with
only $a_{-2}$ oscillators, \ie\ $n_3=n_4=\cdots=0$. If we
consider only their contribution and perform a new Sommerfeld-Watson
transformation on the sum $n_2\rightarrow z_2$, the pole at
$z_2=\half(\alpha(t)+1)$ modifies the power $s^{-1}$ contribution
we found in $A^{(1)}$ to $s^{-1+(\alpha(t)+1)/2}$
$=s^{\alpha(t)/2-1/2}$. This power is exactly the second sister
trajectory $\alpha_2(t)$. Clearly the story now repeats
itself {\it ad infinitum}. As we deform the $z_2$ contour we
can expect to encounter fixed singularities at $z_2=-1$. Their
contribution will be a sum over states of the form
$$a_{-3}^{n_3}a_{-4}^{n_4}\cdots\ket{0,P}\(comp2),$$
associated with a pole factor $1/(-1-2+3n_3+4n_4+\cdots-\alpha(t))$ and
power of $s$, $-1-1+n_3+n_4+\cdots$. The states in \eq{comp2}
with maximal angular momentum are those
with only $n_3\neq0$. Inserting this set of states and continuing
$n_3\rightarrow z_3$ yields a Regge pole at
$z_3=\third(\alpha(t)+1+2)$ and the power of $s$ becomes
$-1-1+n_3=\third\alpha(t)-1$ which is exactly the third
sister $\alpha_3(t)$. At the $k$th step of this process
we generate a singularity
$$z_k={1\over k}(\alpha(t)+1+\cdots+(k-1))=
{1\over k}(\alpha(t)+k(k-1)/2)$$
yielding a power of $s$,
$$-(k-1)+z_k={1\over k}\alpha(t)-{k-1\over2}=\alpha_k(t).$$

Our analysis has shown how the sister trajectories are generated
successively. Each new sister is required to cancel an
unphysical singularity in the residue function of the
previous sister. In this iterative scheme, one could consider
an approximation consisting in deleting the contribution
of all string excitations with mode number larger than some
cutoff $n_C$. Then the first $n_C$ sisters would appear
as trajectories with finite slope, with all the higher ones
replaced by a fixed (double) pole which prevents the included
sister trjectories from dominating at indefinitely large $-t$
(see \Fig{cutoff}).
One can plausibly associate such a fixed power
behavior with a ``hard'' point-like constituent present in the
approximate model. For example,
taking $n_C=1$ gives a crude caricature of the
behavior expected from $QCD$ which does predict hard point-like
constituents. But in string theory, closer examination (\ie\
increasing $n_C$) shows that
this ``hard'' constituent itself has structure associated with
the second modes of excitation. And so it goes, at higher and
higher $-t$ the excitations in the higher modes dominate the
scattering. This correlation of high $-t$ probes with the
high frequency string excitations, shows that one never reveals
truly hard partons. The string is a composite of wee partons
only. It is very interesting that
these high mode contributions enter in just such a way as
to fulfill the Cerulus-Martin bound. That bound is expected
in conventional quantum field theories as a consequence of
unitarity and power boundedness.
Its validity in string theory lends some support
to the idea that the absence of hard partons in string theory
is not inconsistent with physical principles. It is
also suggestive that, in its fundamental formulation,
string theory may well be a new kind of quantum field theory.
\insertfigure{cutoff}{
\centerline{\psfig{figure=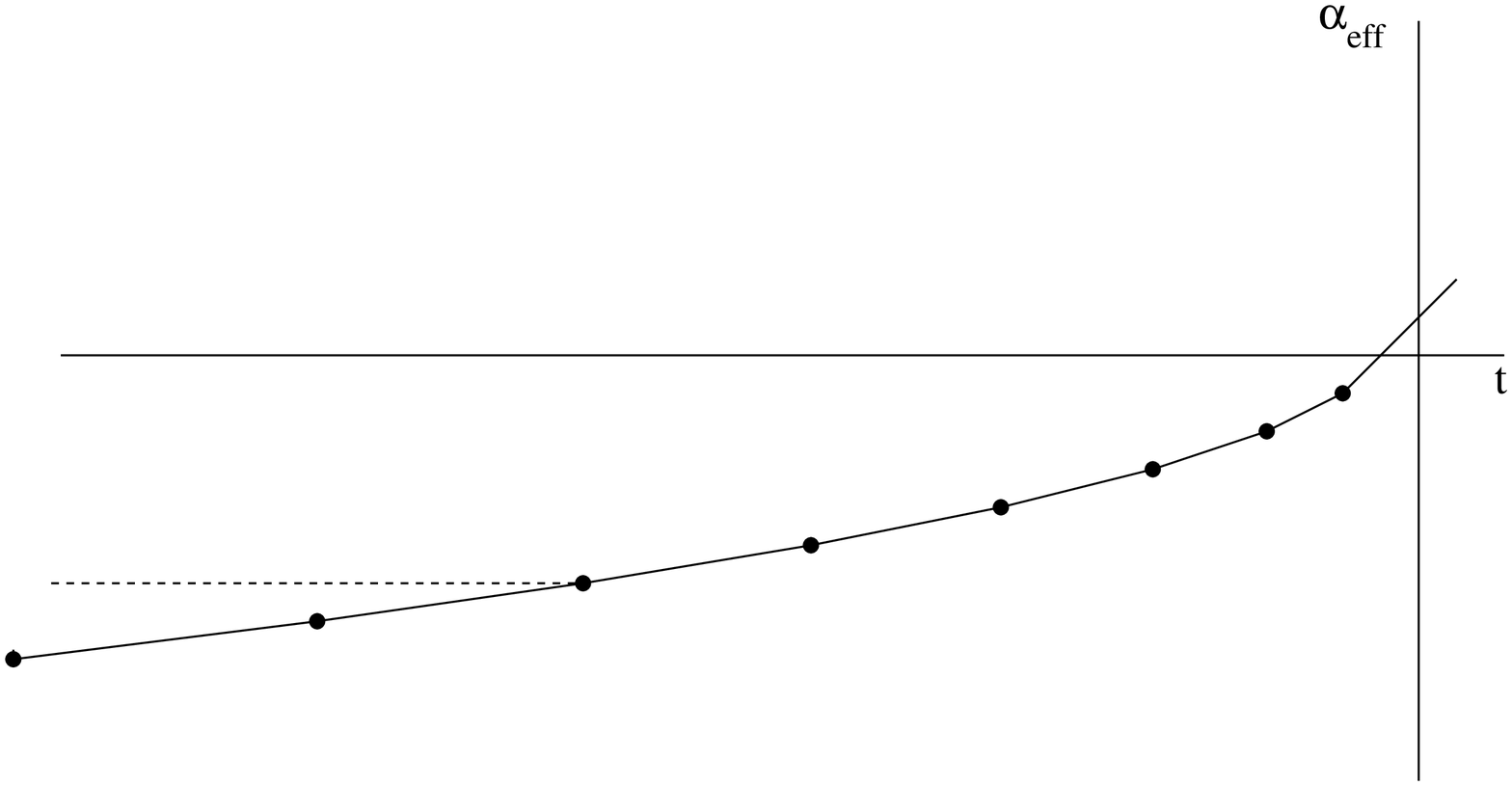,height=5.5cm}}}
{Envelope of the first few sister trajectories. The
dotted line indicates the effective trajectory in a string
model with a mode number cutoff $n_C=6$.}

In conclusion we return to our comparison of string theory and
QCD. Both theories possess a spectrum of particles lying on
Regge trajectories $\alpha(t)$
rising linearly as $t\rightarrow+\infty$.
To avoid conflict with the Cerulus-Martin bound, the leading
trajectory cannot decrease linearly as $t\rightarrow-\infty$.
In QCD this is achieved because that limit is dominated by
hard parton scattering, and consequently the leading trajectory
approaches a constant in this limit. There are no hard partons
in string theory, but the family of sister trajectories envelops
an effective trajectory that saturates the C-M bound.
\endpage
\titlestyle{References}
\reflist{}
\bye